\documentclass{lmcs}
\pdfoutput=1

\usepackage{lastpage}
\lmcsdoi{16}{2}{5}
\lmcsheading{}{\pageref{LastPage}}{}{}%
{May~01,~2019}{May~14,~2020}{}

\usepackage[T1]{fontenc}
\usepackage{amssymb,pifont}
\def \pref#1{\mathbf{pref}(#1)}

\def \fam#1{(#1_{i})_{i\in\mathbb{N}}}

\title[Incomputability of dimension]{On the incomputability of computable
  dimension} 

\author[L. Staiger]{Ludwig Staiger} 
\address{Institut f\"ur Informatik\\ Martin-Luther-Universit\"at\
  Halle-Wittenberg, Germany}
\email{staiger@informatik.uni-halle.de}
\begin{document}

\maketitle

\begin{abstract}
  Using an iterative tree construction we show that for simple computable
  subsets of the Cantor space Hausdorff, constructive and computable
  dimensions might be incomputable.
\end{abstract}
Computable dimension along with constructive dimension was introduced by
Lutz~\cite{siamcomp/Lutz03,ic/Lutz03} as a means for measuring the complexity
of sets of infinite strings ($\omega$-words). Since then and prior to this
constructive and computable dimension were investigated in connection with
Hausdorff dimension (for a detailed account see \cite[Section
13]{book_DowneyHirsch10}). The results of \cite{tocs/Hitch05,ic/St93,tcs/St07}
show that the Hausdorff, constructive and computable dimensions of automaton
definable sets of infinite strings (regular $\omega$-languages) are
computable. In contrast to this Ko \cite{ko98} derived examples of computable
$\omega$-languages with an incomputable Hausdorff dimension.

In this paper we derive examples of computable $\omega$-languages of a simple
structure which have not only incomputable Hausdorff dimension but also
incomputable computable dimension. To this end we use an iteration of finite
trees which resembles the tree construction of Furstenberg
\cite{Furstenberg70} (see also \cite{ciaa/MSS18})

% As a byproduct we obtain simple examples of computable $\omega$-languages
% having incomputable Hausdorff dimension.

Lutz \cite{siamcomp/Lutz03,ic/Lutz03} defines computable and constructive
dimension via $\sigma$-(super)gales. Terwijn \cite{terwijn04,apal/CST06}
observed that this can also be done using Schnorr's concept of combining
martingales with (exponential) order functions \cite[Section
17]{book_Schnorr71}. For the computable $\omega$-languages constructed in this
paper we can show that Schnorr's concept is in some details more precise than
Lutz's approach.

\section{Notation}
\label{sec.nota}

In this section we introduce the notation used throughout the paper.  By
$\mathbb{N} = \{ 0,1,2,\ldots\}$ we denote the set of natural numbers, by
$\mathbb{Q}$ the set of rational numbers, and $\mathbb{R}$ are the real
numbers.

Let $X$ be an alphabet of cardinality $|X|\ge 2$. By $X^*$ we denote the set
of finite words on $X$, including the \textit{empty word} $e$, and
$X^{\omega}$ is the set of infinite strings ($\omega$-words) over $X$.
Subsets of $X^{*}$ will be referred to as \emph{languages} and subsets of
$X^{\omega}$ as \emph{$\omega$-languages}.

For $w\in X^*$ and $\eta\in X^*\cup X^{\omega}$ let $w \cdot{}\eta$ be their
\textit{concatenation}.  This concatenation product extends in an obvious way
to subsets $W \subseteq X^*$ and $B\subseteq X^*\cup X^{\omega}$.  We denote
by $|w|$ the \textit{length} of the word $w\in X^*$ and $\pref B$ is the set
of all finite prefixes of strings in $B\subseteq X^*\cup X^\omega$.

It is sometimes convenient to regard $X^{\omega}$ as Cantor space, that is, as
the product space of the (discrete space) $X$. Here \emph{open} sets in
$X^{\omega}$ are those of the form $W\cdot X^{\omega}$ with $W\subseteq
X^{*}$. \emph{Closed} are sets $F\subseteq X^{\omega}$ which satisfy the
condition $F= \{\xi: \pref\xi\subseteq \pref F\}$.

For a computable domain $\mathcal{D}$, such as $\mathbb{N}$, $\mathbb{Q}$ or
$X^{*}$, we refer to a function $f:\mathcal{D}\rightarrow \mathbb{R}$ as
\emph{left-computable} (or \emph{approximable from below}) provided the set
$\{(d,q): d\in \mathcal{D}\wedge q\in \mathbb{Q}\wedge q< f(d)\}$ is
computably enumerable. Accordingly, a function $f:\mathcal{D}\rightarrow
\mathbb{R}$ is called \emph{right-computable} (or \emph{approximable from
  above}) if the set $\{(d,q): d\in \mathcal{D}\wedge q\in \mathbb{Q}\wedge q>
f(d)\}$ is computably enumerable, and $f$ is \emph{computable} if $f$ is
right- and left-computable.  If we refer to a function
$f:\mathcal{D}\rightarrow \mathbb{Q}$ as computable we usually mean that it
maps the domain $\mathcal D$ to the domain $\mathbb{Q}$, that is, it returns
the exact value $f(d)\in \mathbb{Q}$. If $\mathcal{D}=\mathbb{N}$ we write $f$
as a sequence $\fam q$.

A real number $\alpha\in \mathbb{R}$ is left-computable, right computable or
computable provided the constant function $c_{\alpha}(t)=\alpha$ is
left-computable, right-computable or computable, respectively. $\alpha\in
\mathbb{R}$ is referred to as \emph{computably approximable} if $\alpha=
\lim_{i\to\infty}q_{i}$ for a computable sequence $\fam q$ of rationals. It is
well-known (see e.g. \cite{mlq/ZW01}) that there are left-computable which are
not right-computable and vice versa, and that there are computably
approximable reals which are neither left-computable nor right-computable.

The following approximation property is easily verified.
\begin{pty}\label{pty.approx}
  Let $\fam q$ be a computable family of rationals converging to $\alpha$ and
  let $\fam{q'}, q'_{i}>0,$ be a computable family of rationals converging to
  $0$. If $\alpha$ is not right-computable then there are infinitely many
  $i\in \mathbb{N}$ such that $\alpha - q_{i}> q_{i}'$.  \qed\end{pty} For,
otherwise, $\alpha$ as the limit of $(q_{i}+q'_{i})_{i\in \mathbb{N}}$ would
be right-computable.

\section{Gales and Martingales}
\label{sec.gales}
Hausdorff \cite{hausdorff18} introduced a notion of dimension of a subset $Y$
of a metric space which is now known as its \emph{Hausdorff dimension}, $\dim
Y$; Falconer \cite{book:Falconer03} provides an overview and introduction to
this subject. In the case of the Cantor space $X^{\omega}$, Lutz
\cite{ic/Lutz03} (see also \cite[Section 13.2]{book_DowneyHirsch10}) has found
an equivalent definition of Hausdorff dimension via generalisations of
martingales.

Following Lutz a mapping $d:X^{*}\to [0,\infty)$ will be called an
\emph{$\sigma$-supergale} provided
\begin{equation}
  \label{eq.gales}
  \forall w(w\in X^{*}\to |X|^{\sigma}\cdot d(w)\ge \sum_{x\in X}d(wx))\,.
\end{equation}
A $\sigma$-supergale $d$ is called an $\sigma$-gale if, for all $w\in X^{*}$,
Eq.~(\ref{eq.gales}) is satisfied with
equality. (\emph{Super})\emph{Martingales} are $1$-(super)gales.

From Eq.~(\ref{eq.gales}) one easily infers that if $d,\mathcal{V}: X^{*}\to
[0,\infty)$ satisfy
\begin{equation}
  \label{eq.gales2}
  \forall w(w\in X^{*}\to \frac{\mathcal{V}(w)}{|X|^{(1-\sigma)\cdot|w|}}= d(w))
\end{equation}
then $d$ is a $\sigma$-(super)gale if and only if $\mathcal{V}$ is a
(super)martingale. Thus (super)gales can be viewed as a combination of
(super)martingales with exponential order functions in the sense of Schnorr
\cite[Section 17]{book_Schnorr71} (see also \cite{terwijn04,apal/CST06} or
\cite[Section 13.3]{book_DowneyHirsch10}).

Following Lutz \cite{ic/Lutz03} we define as follows.
\begin{defi}\label{def.HD}\upshape{}
  Let $F\subseteq X^{\omega}$. Then $\alpha$ is the \emph{Hausdorff dimension}
  $\dim F$ of $F$ provided
  \begin{enumerate}[beginpenalty=99]
  \item for all $\sigma> \alpha$ there is a $\sigma$-supergale $d$ such that
    $\forall \xi(\xi\in F\to \limsup\limits_{w\to \xi} d(w)=\infty)$,
    and\footnote{Here $\limsup\limits_{w\to \xi} d(w)$ is an abbreviation for
      $\lim\limits_{n\to\infty} \sup\{d(w): w\in \pref\xi\wedge |w|\ge n\}$.}
  \item for all $\sigma<\alpha$ and all $\sigma$-supergales $d$ it holds
    $\exists \xi(\xi\in F\wedge \limsup\limits_{w\to \xi} d(w)<\infty)$.
  \end{enumerate}
\end{defi}
If the $\omega$-language $F\subseteq X^{\omega}$ is closed in Cantor space and
satisfies a certain balance condition Theorem~4 of \cite{jstplann/St89} shows
that the calculation of its Hausdorff dimension can be simplified. For the
purposes of our investigations the following special case will suffice.
\begin{prop}\label{p.HD}
  Let $F\subseteq X^{\omega}$ be non-empty and satisfy the conditions

  \begin{enumerate}
  \item \rule[-6pt]{0pt}{5pt}$F=\{\xi: \pref\xi\subseteq \pref F\}$ and
  \item $|\pref F\cap w\cdot X^{k}|=|\pref F\cap v\cdot X^{k}|$ for all
    $k\in\mathbb{N}$ and $w,v\in \pref F$ with $|w|=|v|$.
  \end{enumerate}
  Then $\displaystyle\dim F= \liminf\limits_{n\to\infty}\frac{\log_{|X|}|\pref
    F\cap X^{n}|}{n}$\,.\qed
\end{prop}

\section{Iterative Tree Construction}
\label{sec.gen}
The aim of this section is, given a sequence of rationals $\fam q, 0< q_{i}<
1$, to construct an $\omega$-language $F\subseteq X^{\omega}$ with Hausdorff
dimension $\dim F= \liminf_{i\to\infty}q_{i}$ satisfying the conditions (1)
and (2) of Proposition~\ref{p.HD}.
\subsection{Preliminaries}
\label{sec.pre}
As a preparation we show how to find sequences of natural numbers $\fam k$ and
$\fam\ell$ with appropriate properties such that $q_{i}=k_{i}/\ell_{i}$.
\begin{lem}\label{l.gen}
  Let $\fam q$, $0<q_{i}<1$, $q_{i}\ne q_{i+1}$, be a family of positive
  rationals. Then there are families of natural numbers $\fam k$,
  $\fam{\ell}$, $\fam\kappa$, $\fam p$ and $\fam r$, such that
  $q_{i}=k_i/\ell_i$, $\displaystyle q_{i+1}=\frac{r_{i}\cdot
    k_{i}+\kappa_{i}\cdot\ell_{i}}{r_{i}\cdot \ell_{i}+ p_{i}\cdot\ell_{i}}$
  where $\kappa_{i}=\left\{\begin{array}{ll}
      0,&\mbox{ if }q_{i}> q_{i+1}\mbox{ and}\\
      p_{i},&\mbox{ if }q_{i}< q_{i+1}.  \end{array}\right.$ \medskip{}
  \begin{eqnarray}
    \label{eq.gen1}
    \hspace*{-12ex}\mbox{Moreover, for $0\le t\le p_{i}\cdot \ell_{i}$ we have\ }
    q_{i}\ge &\displaystyle\frac{r_{i}\cdot k_{i}}{r_{i}\cdot
      \ell_{i}+t}&\ge q_{i+1},\mbox{ if }q_{i}>q_{i+1}\mbox{ and}\\ \label{eq.gen2}
    q_{i}\le &\displaystyle\frac{r_{i}\cdot k_{i}+t}{r_{i}\cdot
      \ell_{i}+t}&\le q_{i+1},\mbox{ if }q_{i}<q_{i+1}.
  \end{eqnarray}
\end{lem}
\proof
% If $q_{i}=q_{i+1}$ then $r_{i}=1$ and $\kappa_{i}= p_{i}=0$ are sufficient.
Let $q_{i}=k_{i}/\ell_{i}$ and $\displaystyle q_{i+1}=a/b\cdot q_{i}=
\frac{a\cdot k_{i}}{b\cdot \ell_{i}}$, with $a,b\in\mathbb{N}\setminus\{0\},
a\ne b$. Since $1> q_{i+1}$ we have $b\cdot \ell_{i}-a\cdot k_{i}=
a\cdot\frac{q_{i}}{q_{i+1}}\cdot(1-q_{i+1})\cdot\ell_{i}>0$.

Assume $q_{i}>q_{i+1}$.  Then $b>a$ and the equation
\begin{equation}\label{eq.qi+1}
  \frac{r_{i}\cdot k_{i}+\kappa_{i}\cdot\ell_{i}}{r_{i}\cdot
    \ell_{i}+ p_{i}\cdot\ell_{i}}= \frac{a\cdot k_{i}}{b\cdot \ell_{i}}  
\end{equation}
has the solutions $r_{i}=a$, and $p_{i}=(b-a)=a\cdot
(\frac{q_{i}}{q_{i+1}}-1)$ and $\kappa_{i}=0$.

If $q_{i}<q_{i+1}$ then $a> b$ and $r_{i}:= b\cdot \ell_{i}-a\cdot k_{i}=
a\cdot (\frac{q_{i}}{q_{i+1}}\cdot\ell_{i}-k_{i})= a\cdot
q_{i}\cdot(\frac{1}{q_{i+1}}-1)\cdot\ell_{i}$ and
$p_{i}=\kappa_{i}:=(a-b)\cdot k_{i}=a\cdot
q_{i}\cdot(1-\frac{q_{i}}{q_{i+1}})\cdot\ell_{i}$ are solutions of
Eq.~(\ref{eq.qi+1}).

In view of $\kappa_{i}=0$ Eq.~(\ref{eq.gen1}) is obvious. Eq.~(\ref{eq.gen2})
follows inductively from $\frac{k+1}{\ell+1}\ge \frac{k}{\ell}$ whenever $0\le
k< \ell$.  \qed

\bigskip

If the family $\fam q$ is a computable one then the families in
Lemma~\ref{l.gen} can be chosen to be computable. In addition, the values
$\ell_{i}$ and $\ell_{i+1}/\ell_{i}$ can be made arbitrarily large.

\subsection{Tree construction}
\label{sec.tree}
The $\omega$-language $F$ will be the limit of the following sequence of
finite trees $T_{i}$. These trees have a property similar to the one in
Proposition~\ref{p.HD}\,(2) which is referred to as \emph{spherical symmetry}
in \cite{Furstenberg70}.

We define the following auxiliary languages $T_{i}\subseteq X^{\ell_{i}}$ and
$U_{i}\subseteq X^{p_i\cdot \ell_i}$.

Let $T_{0}:= X^{k_{0}}\cdot 0^{\ell_{0}-k_{0}}$ or
$T_{0}:=0^{\ell_{0}-k_{0}}\cdot X^{k_{0}}$ and set
\begin{equation}
  \label{eq.1}
  T_{i+1}:= T_{i}^{r_i}\cdot U_i\mbox{ with }U_{i}:=\left\{
    \begin{array}{ll}
      X^{p_i\cdot \ell_i},&\mbox{ if }q_{i+1}\ge q_{i} \mbox{ and}\\
      \{u_{i}\},&\mbox{ otherwise}
    \end{array}\right.
\end{equation}
\noindent where $u_{i}\in X^{p_i}$ is a fixed word. Then $\ell_{i+1}=
(r_{i}+p_{i})\cdot\ell_{i}$. Thus $T_{i+1}$ consists of a concatenation of
$r_{i}$ copies of $T_{i}$ plus an appendix $U_{i}$ of length $p_{i}\cdot
\ell_{i}$. The values $r_{i}$ and $p_{i}$ are referred to as repetition or
prolongation factors, respectively.

By induction one proves \begin{equation}
  \label{eq.2}|T_{i}|= |X|^{q_{i}\cdot \ell_{i}}\,.
\end{equation}
\begin{pty}\label{pty.1}The trees $T_{i}$ have the following properties.
  Let $\ell\le \ell_{i}$.
  \begin{enumerate}
  \item Prefix property:\rule[-6pt]{0pt}{5pt} $\pref{T_{i+1}}=
    \bigcup_{j=0}^{r_{i}-1}T_{i}^{j}\cdot \pref{T_{i}} \cup T_{i}^{r_{i}}\cdot
    \pref{U_{i}}$, \label{it.1}
  \item Extension property:\rule[-6pt]{0pt}{5pt} $\pref{T_{i}}\cap X^{\ell}=
    \pref{T_{i+1}}\cap X^{\ell}$, and\label{it.2}
  \item Spherical symmetry:\rule[-6pt]{0pt}{5pt} $\pref{T_{i}}\cap X^{\ell}=
    (\pref{T_{i}}\cap X^{\ell-1})\cdot X$ or\label{it.3}
  \item[]\hspace{20.3ex} $|\pref{T_{i}}\cap X^{\ell}|= |\pref{T_{i}}\cap
    X^{\ell-1}|.$
    \qed
  \end{enumerate}
  \end{pty}

\subsection{The infinite tree}
\label{sec.infinit}
We define our $\omega$-language $F$ having the properties mentioned in
Proposition~\ref{p.HD} as $F:= \bigcap_{i\in\mathbb{N}}T_{i}\cdot X^{\omega}$
where the family $\fam T$ satisfies Eq.~(\ref{eq.1}).

Before we proceed to further properties of $\fam T$ and $F$ we mention a
general property.
\begin{lem}\label{l.schnitt}
  Let $T_{i}\subseteq X^{*}$, $T_{i+1}\subseteq T_{i}\cdot X\cdot X^{*}$,
  $T_{i}\subseteq \pref{T_{i+1}}$ and $F:= \bigcap_{i\in\mathbb{N}}T_{i}\cdot
  X^{\omega}$. Then $\pref F= \bigcup_{i\in\mathbb{N}}\pref{T_{i}}$.

  If, moreover, all $T_{i}$ are finite then $F:= \{\xi:\xi\in X^{\omega}\wedge
  \pref\xi\subseteq \bigcup_{i\in\mathbb{N}}\pref{T_{i}}\}$.
\end{lem}
\proof In view of $T_{i+1}\subseteq T_{i}\cdot X\cdot X^{*}$ we have
$T_{i+1}\cdot X^\omega \subseteq T_{i}\cdot X^\omega$ and also $|w|\ge i$ for
$w\in T_{i}$.

If $w\in \pref F$ then $w\in \pref\xi$ where $\xi\in F\subseteq T_{i}\cdot
X^{\omega}$ for $i>|w|$. Consequently, $w\in \pref{T_{i}}$.

Using the condition $T_{i}\subseteq \pref{T_{i+1}}$, by induction we obtain
that for every $w\in\pref{T_{i}}$ there is an infinite chain $(w_{j})_{j\ge
  i}$ such that $w_{j}\in T_{j}$ and $w\sqsubseteq w_{i}\sqsubset
w_{i+1}\sqsubset \cdots$. Thus there is a $\xi \in F$ with $w\sqsubset\xi$.

If the languages $T_{i}$ are finite $F=\bigcap_{i\in\mathbb{N}}T_{i}\cdot
X^{\omega}$ is closed in the product topology of the space $X^{\omega}$ which
implies $F:= \{\xi:\xi\in X^{\omega}\wedge \pref\xi\subseteq\pref{F}\}$.  \qed

\bigskip

Lemma~\ref{l.schnitt} shows that $F:= \{\xi:\xi\in X^{\omega}\wedge
\pref\xi\subseteq \bigcup_{i\in\mathbb{N}}\pref{T_{i}}\}$ for the family
$\fam{T}$ defined in Section~\ref{sec.tree}.

From the spherical symmetry of $T_{i}$ (see
Property~\ref{pty.1}\,(\ref{it.3})) the $\omega$-language
$F=\bigcap_{i\in\mathbb{N}}T_{i}\cdot X^{\omega}$ inherits the following
balance property of Proposition~\ref{p.HD}\,(2).
\begin{lem}\label{l.balance}Let $F=\bigcap_{i\in\mathbb{N}}T_{i}\cdot
  X^{\omega}$ where the $T_{i}$ are defined by Eq.~(\ref{eq.1}).  Then for all
  $k\in\mathbb{N}$ and $w,v\in \pref F$ with $|w|=|v|$ we have
  \[|w\cdot X^{k}\cap \pref F|=|v\cdot X^{k}\cap \pref F|\,.\]
\end{lem}
\proof We proceed by induction on $k$. Let $k=1$. Then for all $w,v\in \pref
F$ with $|w|=|v|$ either $\pref{F}\cap X^{|u|+1}= (\pref{F}\cap X^{|u|})\cdot
X$ or $|\pref{F}\cap X^{|u|+1}|= |\pref{F}\cap X^{|u|}|$ ($u\in \{w,v\}$).

In the first case we have $|w\cdot X\cap \pref F|=|X|=|v\cdot X\cap \pref F|$
and in the second $|w\cdot X\cap \pref F|=1=|v\cdot X\cap \pref F|$.

Let the assertion be proved for $k$ and all pairs $u,u'\in \pref F$ of the
same length. Let $w,v\in \pref F$ with $|w|=|v|$ and consider words $w',v'\in
X^k$ such that $w\cdot w',v\cdot v'\in \pref F$. Then from the spherical
symmetry we obtain either $\pref{F}\cap X^{|u|+1}= (\pref{F}\cap X^{|u|})\cdot
X$ or $|\pref{F}\cap X^{|u|+1}|= |\pref{F}\cap X^{|u|}|$ for $u\in \{w\cdot
w',v\cdot v'\}$ and we proceed as above.

Since, by our assumption $|\{w': |w'|=k\wedge w\cdot w'\in \pref F\}|=|\{v':
|v'|=k\wedge v\cdot v'\in \pref F\}|$, the assertion follows.  \qed

\bigskip As a consequence of Lemmas~\ref{l.schnitt}, \ref{l.balance} and
Proposition~\ref{p.HD} we obtain the following.
\begin{cor}\label{c.HD}Let $F=\bigcap_{i\in\mathbb{N}}T_{i}\cdot X^{\omega}$
  where the $T_{i}$ are defined by Eq.~(\ref{eq.1}).  Then $\dim F=
  \liminf\limits_{n\to\infty}\frac{\log_{|X|}|\pref F\cap X^{n}|}{n}$.
  \qed\end{cor}

\bigskip{}

Next we investigate in more detail the structure function
$s_{F}:\mathbb{N}\to\mathbb{N}$ where $s_F(\ell):= |\pref F\cap X^{\ell}|$.
First, Lemma~\ref{l.schnitt} implies
\begin{equation}
  \label{eq.schnitt}
  \pref F\cap X^{\ell}=\pref{T_{i}}\cap
  X^{\ell}\mbox{ whenever }  \ell\le\ell_{i}\,.
\end{equation}
From Eqs.~(\ref{eq.1}) and (\ref{eq.2}) and the properties of the tree family
$(T_{i})_{i\in\mathbb{N}}$ we obtain for the intervals $\ell_{i}\le \ell\le
\ell_{i+1}$:
\begin{lem}\label{l.T1}Let $F=\bigcap_{i\in\mathbb{N}}T_{i}\cdot X^{\omega}$
  where the $T_{i}$ are defined by Eq.~(\ref{eq.1}). Then the structure
  function $s_{F}:\mathbb{N}\to \mathbb{N}$ satisfies the following relations.
  \begin{enumerate}
  \item In the interval $[j\cdot \ell_{i}, (j+1)\cdot\ell_{i}]$ where
    $j< r_{i}$:\\
    \[s_{F}(j\cdot\ell_{i}+t) =s_{F}(\ell_{i})^{j}\cdot s_{F}(t) \mbox{ for }
    0\le t\le \ell_{i}\]\label{l.T11}
  \item In the subinterval $[j\cdot \ell_{i}+j'\cdot\ell_{i-1},j\cdot
    \ell_{i}+(j'+1)\cdot\ell_{i-1}]$ where $j'< r_{i-1}$:\\
    \[s_{F}(j\cdot\ell_{i}+j'\cdot \ell_{i-1}+t) =s_{F}(\ell_{i})^{j}\cdot
    s_{F}(\ell_{i-1})^{j'}\cdot s_{F}(t)\mbox{ for }0\le
    t<\ell_{i-1}\,.\]\label{l.T12}
  \item In the interval $[r_{i}\cdot \ell_{i}, \ell_{i+1}]$:
    \[s_{F}(r_{i}\cdot\ell_{i}+t) =\left\{
      \begin{array}{lll}
        s_{F}(\ell_{i})^{r_{i}},&\mbox{if}&|U_{i}|=1\mbox{ and}\\[3pt]
        s_{F}(\ell_{i})^{r_{i}}\cdot |X|^{t},&\mbox{if}&U_{i}= X^{p_{i}\cdot
          \ell_{i}}
      \end{array}\right.\mbox{ for }0\le t\le p_{i}\cdot\ell_{i}\,.
    \tag*{\qed}
  \]\label{l.T13}
  \end{enumerate}
\end{lem}

This yields the following connection to the values $q_{i}$.  In order to
connect our considerations to the application of Proposition~\ref{p.HD} we
consider the values of $\frac{\log_{|X|} s_{F}(n)}{n}$ instead of $s_{F}(n)$.

From Eqs.~(\ref{eq.schnitt}) and (\ref{eq.2}) we obtain
\begin{equation}
  \label{eq.sF0}
  \frac{\log_{|X|} s_{F}(j\cdot\ell_{i})}{j\cdot\ell_{i}} =q_{i}\,.
\end{equation}  
Now we use the identities of Lemma~\ref{l.T1} and Eqs.~(\ref{eq.gen1}) and
(\ref{eq.gen2}) to bound $\frac{\log_{|X|} s_{F}(\ell)}{\ell}$ in the range
$\ell_{i}\le \ell\le \ell_{i+1}= r_{i}\cdot\ell_{i}+n_{i}\cdot\ell_{i}$.

For $\ell_{i}\le \ell<r_{i}\cdot \ell_{i}$ we have $\ell=j\cdot \ell_{i}+
j'\cdot\ell_{i-1} +t$ where $0\le t< \ell_{i-1}$, and
Lemma~\ref{l.T1}\,(\ref{l.T11}) and (\ref{l.T12}) yield
\begin{eqnarray}\nonumber
  \frac{\log_{|X|} s_{F}(\ell)}{\ell}&\ge& \frac{j\cdot \ell_{i}}{\ell}\cdot
  q_{i}+ \frac{j'\cdot \ell_{i-1}}{\ell}\cdot q_{i-1}\\\label{eq.sF1}%\nonumber
  &\ge& \frac{j\cdot \ell_{i}+j'\cdot  \ell_{i-1}}{\ell}\cdot  \min\{q_{i-1},q_{i}\}\\\nonumber
  & \ge& (1- \frac{\ell_{i-1}}{\ell_{i}})\cdot  \min\{q_{i-1},q_{i}\}
\end{eqnarray}
If $r_{i}\cdot \ell_{i}\le \ell\le \ell_{i+1}$, that is, for $\ell=r_{i}\cdot
\ell_{i}+t$ where $t\le \ell_{i+1}-r_{i}\cdot \ell_{i}$, following
Eqs.~(\ref{eq.gen1}) and (\ref{eq.gen2}), respectively, we have according to
Lemma~\ref{l.T1}\,(\ref{l.T13})
\begin{eqnarray}
  \label{eq.sF2}
  q_{i}\ge\frac{\log_{|X|} s_{F}(\ell)}{\ell}&=& \frac{\log_{|X|} s_{F}(r_{i}\cdot
    \ell_{i})}{r_{i}\cdot \ell_{i}+t}\ge q_{i+1}\mbox{ if }q_{i}>q_{i+1}\\\label{eq.sF3}
  q_{i}\le\frac{\log_{|X|} s_{F}(\ell)}{\ell}&=& \frac{\log_{|X|}
    s_{F}(r_{i}\cdot \ell_{i})+t}{r_{i}\cdot \ell_{i}+t}\le q_{i+1}\mbox{ if }q_{i}<q_{i+1}
\end{eqnarray}
The considerations in Eqs.~(\ref{eq.sF0}), (\ref{eq.sF1}), (\ref{eq.sF2}) and
(\ref{eq.sF3}) show the following.
\begin{lem}\label{l.liminf}
  If the sequence $\fam{\ell}$ is chosen in such a way that
  $\liminf\limits_{i\to\infty}\frac{\ell_{i-1}}{\ell_{i}}=0$ then
  \[\textstyle\liminf\limits_{\ell\to\infty}\frac{\log_{|X|}
    s_{F}(\ell)}{\ell}= \liminf\limits_{i\to\infty}q_{i}\,.\]
\end{lem}
\proof In view of Eq.~(\ref{eq.sF0}) the limit cannot exceed
$\liminf\limits_{i\to\infty}q_{i}$.

On the other hand, by Eqs.~(\ref{eq.sF1}), (\ref{eq.sF2}) and (\ref{eq.sF3}),
for $\ell_{i}\le \ell\le \ell_{i+1}$, the intermediate values satisfy
$\frac{\log_{|X|} s_{F}(\ell)}{\ell} \ge (1- \frac{\ell_{i-1}}{\ell_{i}})\cdot
\min\{q_{i-1},q_{i},q_{i+1}\}$.  \qed

\subsection{Monotone families $\fam{q}$}
If the sequence $\fam{q}$ is monotone we can simplify the above considerations
of Eq.~(\ref{eq.sF1}).
\begin{prop}
  Let the sequence $\fam{q}$ be monotone and
  $\lim_{i\to\infty}q_{i}=\alpha$.\label{p.mono}
  \begin{enumerate}
  \item If $\fam{q}$ is decreasing and $T_{0}= X^{k_{0}}\cdot
    0^{\ell_{0}-k_{0}}$ then $s_{F}(\ell)\ge |X|^{\alpha\cdot \ell}$, for all
    $\ell\in\mathbb{N}$.\label{p.mono1}
  \item If $\fam{q}$ is increasing and $T_{0}= 0^{\ell_{0}-k_{0}}\cdot
    X^{k_{0}}$ then $s_{F}(\ell)\le |X|^{\alpha\cdot \ell}$, for all
    $\ell\in\mathbb{N}$.\label{p.mono2}
  \end{enumerate}
\end{prop}
\proof If $\fam{q}$ is decreasing we start with $T_{0}= X^{k_{0}}\cdot
0^{\ell_{0}-k_{0}}$ and have $s_{F}(\ell)\ge|X|^{q_{0}\cdot \ell}\ge
|X|^{\alpha\cdot \ell}$ for $\ell\le\ell_{0}$. Then we use
Eqs.~(\ref{eq.schnitt}) and (\ref{eq.1}) and proceed by induction.

$s_{F}(j\cdot \ell_{i}+t)= s_{F}(j\cdot \ell_{i})\cdot
s_{F}(t)\ge|X|^{q_{i}\cdot \ell_{i}}\cdot |X|^{\alpha\cdot t} \ge
|X|^{\alpha\cdot \ell}$ for $j<r_{i}$.  In the range $r_{i}\cdot
\ell_{i}\le\ell\le\ell_{i+1}$ we have according to Eq.~(\ref{eq.sF2})
$s_{F}(\ell)\ge |X|^{q_{i+1}\cdot \ell}\ge |X|^{\alpha\cdot \ell}$.

If $\fam{q}$ is increasing we start with $T_{0}= 0^{\ell_{0}-k_{0}}\cdot
X^{k_{0}}$ and have $s_{F}(\ell)\ge|X|^{q_{0}\cdot \ell}\le |X|^{\alpha\cdot
  \ell}$ for $\ell\le\ell_{0}$. Again we use Eqs.~(\ref{eq.schnitt}) and
(\ref{eq.1}) and proceed by induction.

$s_{F}(j\cdot \ell_{i}+t)= s_{F}(j\cdot \ell_{i})\cdot
s_{F}(t)\le|X|^{q_{i}\cdot \ell_{i}}\cdot |X|^{\alpha\cdot t} \le
|X|^{\alpha\cdot \ell}$ for $j<r_{i}$.  In the range $r_{i}\cdot
\ell_{i}\le\ell\le\ell_{i+1}$ we have according to Eq.~(\ref{eq.sF3})
$s_{F}(\ell)\le |X|^{q_{i+1}\cdot \ell}\le |X|^{\alpha\cdot \ell}$.  \qed

\section{Incomputable dimensions}
\label{sec.ex}
\subsection{Hausdorff dimension}\label{sec.HD}

In this section we provide the announced examples. First we have the
following.
\begin{lem}\label{l.Fcomp}
  If the sequence $\fam q$ of rationals $0<q_{i}<1, q_{i}\ne q_{i+1},$ is
  computable then one can construct an $\omega$-language $F\subseteq
  X^{\omega}$ according to the tree construction such that $\pref F$ is a
  computable language.
\end{lem}
\proof Construct from $\fam q$ the numerator and denominator sequences $\fam
k$ and $\fam\ell$ and the corresponding sequences for the repetition and
prolongation factors $\fam r$ and $\fam p$. Then in view of Eq.~(\ref{eq.1})
the assertion is obvious.  \qed

Our lemma shows that the $\omega$-language $F\subseteq X^{\omega}$ has a very
simple computable structure (compare with \cite[Section 4.2]{tcs/St07}).

Next we show that the Hausdorff dimension of a computable $\omega$-language
$F\subseteq X^{\omega}$ as in Lemma~\ref{l.Fcomp} may be incomputable.
\begin{thm}\label{th.Fcomp1}
  If the sequence $\fam q$ of rationals $0<q_{i}<1, q_{i}\ne q_{i+1},$ is
  computable and $\alpha=\liminf_{i\to\infty}q_{i}$ then there is an
  $\omega$-language $F\subseteq X^{\omega}$ such that $\pref F$ is a
  computable language and $\dim F=\alpha$.
\end{thm}
\proof Construct from $\fam q$ the numerator and denominator sequences $\fam
k$ and $\fam\ell$ such that
$\liminf_{i\to\infty}\frac{\ell_{i}}{\ell_{i+1}}=0$. Then the assertion
follows from Lemmas~\ref{l.liminf}, \ref{l.Fcomp} and Corollary~\ref{c.HD}.
\qed

Theorem 3.4 of \cite{ko98} proves a similar result where the achieved
Hausdorff dimension $\alpha$ is a computably approximable number. In
\cite{mlq/ZW01} it is shown that there are reals which are not computably
approximable of the form $\liminf_{i\to\infty} q_{i}$ where $\fam q$ is a
computable sequence.

\subsection{Computable dimension}\label{sec.martin}
If we require the supergales in Definition~\ref{def.HD} to be computable
mappings we obtain the definition of computable dimension
$\dim_{\mathrm{comp}} F$ of \cite{tocs/Hitch05,ic/Lutz03}.  In view of
Eq.~(\ref{eq.gales2}) we may, as in Section~13.15 of
\cite{book_DowneyHirsch10}, define the computable dimension of an
$\omega$-language $E\subseteq X^{\omega}$ via martingales.

\begin{defi}\label{def.compD}\upshape{}
  Let $F\subseteq X^{\omega}$. Then $\alpha$ is the \emph{computable
    dimension} of $F$ provided
  \begin{enumerate}
  \item for all $\sigma> \alpha$ there is a computable martingale
    $\mathcal{V}$ such that\\ $\forall \xi(\xi\in F\to \limsup\limits_{w\to
      \xi} \frac{\mathcal{V}(w)}{|X|^{(1-\sigma)\cdot|w|}}=\infty)$, and
  \item for all $\sigma<\alpha$ and all computable martingales $\mathcal{V}$
    it holds\\$\exists \xi(\xi\in F\wedge \limsup\limits_{w\to
      \xi}\frac{\mathcal{V}(w)}{|X|^{(1-\sigma)\cdot|w|}}<\infty)$.
  \end{enumerate}
\end{defi}
The inequality $\dim F\le \dim_{\mathrm{comp}}F$ is immediate.

\vbox{We associate with every non-empty $\omega$-language $E\subseteq
  X^{\omega}$ a martingale $\mathcal{V}_{E}$ in the following way.
  \begin{defi}\label{def.martin}
    \begin{eqnarray*}
      \mathcal{V}_{E}(e)&:=& 1\\
      \mathcal{V}_{E}(wx)&:=&\left\{
        \begin{array}{ll}
          \frac{|X|}{|\pref E \cap
            w\cdot X|}\cdot\mathcal{V}_{E}(w),
          &\mbox{ if }x\in X \mbox{ and }wx\in \pref E\mbox{, and}\\
          0,&\mbox{ otherwise.}
        \end{array}\right.
    \end{eqnarray*}
  \end{defi}} In view of the spherical symmetry, for $F$ defined as in
Section~\ref{sec.infinit}, we obtain
\begin{equation}\label{eq.VF}
  \mathcal{V}_{F}(w)= |X|^{|w|}/s_{F}(|w|), \mbox{ if }w\in \pref F\,.
\end{equation}
Moreover, if $\pref F$ is computable then $s_{F}$ and $\mathcal{V}_{F}$ are
computable mappings.
\begin{thm}\label{th.compD}
  If the sequence $\fam q$ of rationals $0<q_{i}<1, q_{i}\ne q_{i+1},$ is
  computable and $\alpha=\liminf_{i\to\infty}q_{i}$ then there is an
  $\omega$-language $F\subseteq X^{\omega}$ such that $\pref F$ is a
  computable language and $\dim F=\dim_{\mathrm{comp}} F=\alpha$.
\end{thm}
\proof We use the $\omega$-language $F$ defined in the proof of
Theorem~\ref{th.Fcomp1} and the associated computable martingale
$\mathcal{V}_{F}$.

Let $\sigma>\alpha=\liminf_{i\to\infty}q_{i}$. Then
$(\sigma-q_{i})>(\sigma-\alpha)/2 >0$ for infinitely many
$i\in\mathbb{N}$. Since $s_{F}(\ell_{i})=|X|^{q_{i}\cdot \ell_{i}}$ (see
Eq.~(\ref{eq.sF0})), we have $\mathcal{V}_{F}(w)/|X|^{(1-\sigma)\cdot|w|}=
|X|^{(\sigma-q_{i})}\ge |X|^{(\sigma-\alpha)/2}$ for $w\in \pref F\cap
X^{\ell_{i}}$. This shows
$\limsup_{w\to\xi}\mathcal{V}_{F}(w)/|X|^{(1-\sigma)\cdot|w|}=\infty$ for all
$\xi\in F$, that is, $\dim_{\mathrm{comp}} F\le \alpha$.

The other inequality follows from $\dim F\le \dim_{\mathrm{comp}} F$ and
Theorem~\ref{th.Fcomp1}.  \qed

\bigskip{}

In certain cases we can achieve even the borderline value
\begin{equation}
  \label{eq.VF1}
  \limsup_{w\to\xi}\frac{\mathcal{V}_{F}(w)}{|X|^{(1-\dim
      F)\cdot|w|}}=\limsup_{n\to\infty}\frac{|X|^{\dim F\cdot
      n}}{s_{F}(n)}=\infty\mbox{ for all }\xi\in F\,.
\end{equation}
\begin{thm}\label{th.Fcomp2}
  Let $\fam q, 0<q_{i}<1, q_{i}\ne q_{i+1},$ be a computable sequence of
  rationals with $\liminf_{i\to\infty}q_{i}=\alpha$. If $\alpha$ is not
  right-computable then there is an $\omega$-language $F\subseteq X^{\omega}$
  such that $\pref F$ is a computable language, $\dim F=\dim_{\mathrm{comp}}
  F=\alpha$ and Eq.~(\ref{eq.VF1}) is satisfied.
\end{thm}
\proof We construct $F$ as in the proof of Theorem~\ref{th.Fcomp1} requiring
additionally that $\ell_{i}\ge i^{2}$. Then $\pref F$ is computable and $\dim
F=\dim_{\mathrm{comp}} F=\alpha$. In view of Property~\ref{pty.approx} there
are infinitely many $i\in \mathbb{N}$ with $\alpha - \frac{1}{i}>q_{i}$ and,
consequently, $s_{F}(\ell_{i})=|X|^{q_{i}\cdot \ell_{i}}\le |X|^{\alpha\cdot
  \ell_{i}-\ell_{i}/i}$. This shows
$\limsup\limits_{n\to\infty}\frac{|X|^{\alpha\cdot n}}{s_{F}(n)}\ge
\limsup\limits_{i\to\infty}|X|^{\ell_{i}/i}=\infty$.  \qed

\subsection{Comparison of gales and martingales}\label{sec.compar}
In this final part we compare the precision with which (super)gales and
martingales achieve the value of computable dimension of a subset $E\subseteq
X^{\omega}$. In Theorem~\ref{th.Fcomp2} we have seen that there are
$\omega$-languages $F\subseteq X^{\omega}$ such that $\dim_{\mathrm{comp}}
F=\alpha$ and $\limsup_{w\to\xi}
\mathcal{V}_{F}(w)/|X|^{(1-\alpha)\cdot|w|}=\infty$ for all $\xi\in F$, that
is, the computable martingale $\mathcal{V}_{F}$ ``matches'' exactly the value
of the computable dimension of $F$. The following theorem shows that this is,
in some cases, not possible for supergales.

First, observe that, for $\sigma'\ge \sigma$ any $\sigma$-supergale
$d:X^{*}\to [0,\infty)$ is also a $\sigma'$-supergale. Thus computable
$\sigma$-supergales exist for all $\sigma\in[0,1]$.

We define the \emph{cut point} $\chi_{d}$ of a supergale $d$ as the smallest
value $\sigma$ for which $d$ can be an $\sigma$-supergale.
\begin{equation}
  \label{eq.cut}
  \chi_{d}:= \inf\left\{\sigma: \forall w\left(|X|^{\sigma}\cdot d(w)\ge
      \sum\nolimits_{x\in X}d(wx)\right)\right\}\,.
\end{equation}
If $d$ is a computable mapping then $\chi_{d}$ as
$\sup\{q:q\in\mathbb{Q}\wedge \exists w(|X|^q\cdot d(w)< \sum\nolimits_{x\in
  X}d(wx)) \}$ is a left-computable real number. For computable $\sigma$-gales
$d$ the cut point $\chi_{d}$ coincides with $\sigma$ and is necessarily a
computable real.
\begin{thm}\label{th.match}Let $\fam q, 0<q_{i}<1, q_{i}\ne q_{i+1},$ be a
  computable sequence of rationals with $\liminf_{i\to\infty}q_{i}=\alpha$.
  If $\alpha$ is neither left- nor right-computable then there is an
  $\omega$-language $F\subseteq X^{\omega}$ such that $\pref F$ is a
  computable language, $\alpha=\dim F=\dim_{\mathrm{comp}} F$,
  Eq.~(\ref{eq.VF1}) is satisfied but there is no computable $\alpha$-supergale
  with $\limsup_{w\to\xi}d(w)=\infty$ for all $\xi\in F$.
\end{thm}
\proof In view of the preceding Theorems~\ref{th.Fcomp1} and \ref{th.Fcomp2}
it suffices to show that under the additional assumption that $\alpha$ is not
left-computable no computable $\alpha$-supergale satisfies
$\limsup_{w\to\xi}d(w)=\infty$ for all $\xi\in F$.

Assume the contrary. Since $\alpha$ is not left-computable, the cut point
$\chi_{d}$ of the computable $\alpha$-supergale $d$ cannot coincide with
$\alpha$. Hence $\alpha>\chi_{d}$, and we have some rational number $q,
\alpha>q>\chi_{d}$. Consequently, $d$ is a $q$-supergale with
$\limsup_{w\to\xi}d(w)=\infty$ for all $\xi\in F$. This contradicts
$q<\alpha=\dim_{\mathrm{comp}} F$ \qed

\bigskip{}

Since there are computably approximable reals which are neither right- not
left-computable Theorem~\ref{th.match} shows that in some cases Schnorr's
\cite{book_Schnorr71} combination of martingales with (exponential) order
functions can be more precise than Lutz's approach via supergales.

\section{Concluding remark}
\label{sec.conc}

As the constructive dimension of subsets of $X^{\omega}$ is sandwiched between
the computable and the Hausdorff dimension
(\cite{siamcomp/Lutz03,ic/Lutz03,tocs/Hitch05}) the result of
Theorem~\ref{th.compD} holds likewise for constructive dimension, too.

\section*{Acknowledgement}
I wish to thank the anonymous reviewers for their suggestions for improving
the presentation of this article.

\bibliographystyle{alpha}
\bibliography{/home/staiger/texmf/eigen,/home/staiger/texmf/hausdorff,/home/staiger/texmf/lcr2010}
% \begin{thebibliography}{alpha}
  
% \end{thebibliography}
\end{document}